%% file: root.tex
\documentclass[a4paper, 10pt, conference]{ieeeconf}      

\IEEEoverridecommandlockouts                              

\usepackage{fainekos-macros}
\usepackage{abbr_l2f}
\usepackage{graphics} 
\usepackage{amsmath} 
\usepackage{amssymb}  
\usepackage{xcolor}
\usepackage{graphicx}
\usepackage{multirow}
\usepackage{subcaption}
\usepackage{algorithm}
\usepackage[algo2e]{algorithm2e}
\usepackage[breaklinks=true, colorlinks=false,linkcolor=black, bookmarks=true, citecolor=black, urlcolor=black,linktoc=all]{hyperref}

\newtheorem{mydef}{Definition}
\newtheorem{myass}{Assumption}
\newtheorem{myprob}{Problem}

\newtheorem{corollary}{Corollary}[theorem]

\title{\LARGE \bf
Learning-to-Fly: Learning-based Collision Avoidance for Scalable Urban Air Mobility
}

\author{Al\"ena Rodionova$^{1,^*}$, Yash Vardhan Pant$^{2,^*}$, Kuk Jang$^{1}$, Houssam Abbas$^{3}$, Rahul Mangharam$^{1}$
\thanks{$^{1}$Department of Electrical and Systems Engineering, University of Pennsylvania, Philadelphia, PA, USA
	{\tt\footnotesize \{alena.rodionova, jangkj, rahulm\}@seas.upenn.edu}.}%
\thanks{$^{2}${Department of Electrical Engineering and Computer Sciences, University of California, Berkeley, USA}
	{\tt\footnotesize yashpant@berkeley.edu}.}%
\thanks{$^{3}$School of Electrical Engineering and Computer Science, Oregon State University, Corvallis, USA
	{\tt\footnotesize houssam.abbas@oregonstate.edu}.}%
\thanks{$^*$The authors contributed equally.}
}%

\begin{document}
\maketitle
\thispagestyle{empty}
\pagestyle{empty}

\input{chapters/abstract}
\input{chapters/introduction_new}

\input{chapters/preliminaries}
\input{chapters/problemStatement}
\input{chapters/experimental}

\input{chapters/casestudy}
\input{chapters/futurework}

\addtolength{\textheight}{-2cm}


\bibliographystyle{IEEEtran}
\bibliography{fbl_l2f_refs}
\end{document}

%% file: chapters/abstract.tex
\begin{abstract}
With increasing urban population, there is global interest in Urban Air Mobility (UAM), where hundreds of autonomous Unmanned Aircraft Systems (UAS) execute missions in the airspace above cities. Unlike traditional human-in-the-loop air traffic management, UAM requires decentralized autonomous approaches that scale for an order of magnitude higher aircraft densities and are applicable to urban settings. We present Learning-to-Fly (L2F), a decentralized on-demand airborne collision avoidance framework for multiple UAS that allows them to independently plan and safely execute missions with spatial, temporal and reactive objectives expressed using Signal Temporal Logic. We formulate the problem of predictively avoiding collisions between two UAS without violating mission objectives as a Mixed Integer Linear Program (MILP). This however is intractable to solve online. Instead, we develop L2F, a two-stage collision avoidance method that consists of: 1) a learning-based decision-making scheme and 2) a distributed, linear programming-based UAS control algorithm. Through extensive simulations, we show the real-time applicability of our method which is $\approx\!6000\times$ faster than the MILP approach and can resolve $100\%$ of collisions when there is ample room to maneuver, and shows graceful degradation in performance otherwise. We also compare L2F to two other methods and demonstrate an implementation on quad-rotor robots.
\end{abstract}

%% file: chapters/introduction_new.tex
\section{Introduction}


The development of safe and reliable UAS Traffic Management (UTM) is necessary to enable Urban Air Mobility (UAM) \cite{NAP25646}. The two fundamental issues here are: a) mission planning for UAS fleets with guarantees on 
safety and performance, and b) real-time airborne collision avoidance (CA) methods so UAS run by different operators can share the airspace without a priori approval of all flight plans. Tackling the planning and inter-UAS collision avoidance jointly yields a computationally intractable problem as the number of UAS in the airspace increase \cite{pant2018fly, SahaRSJ14}. So we separate these two aspects in a manner where individual UAS (or those in the same fleet) plan independently, which in turn requires an approach for runtime collision avoidance. The scalability of this will be essential in UAM applications as there will be no central authority to monitor and enforce UAS safety for hundreds of drones per square mile. This stands in contrast to the existing Air-Traffic Control and collision avoidance methods for commercial aviation, like TCAS-II, which was designed to operate in traffic densities of up to 0.3 aircraft per square nautical mile (nmi), i.e., 24 aircraft within a 5 nautical mile radius, which was the highest traffic density envisioned over the next 20 years \cite{TCAS}.

Airborne collision avoidance however is a complex problem. With high-speed UAS operating at low altitudes in cluttered urban airspace, decisions for collision avoidance need to be made within fractions of a second. The CA system must also be able to take into account the environment (e.g. buildings and other infrastructure, altitude limits, geofenced areas etc.) around it, making the problem harder than simply avoiding inter-UAS collisions. 

\begin{figure}[t]
	\centering
\includegraphics[width=0.35\textwidth]{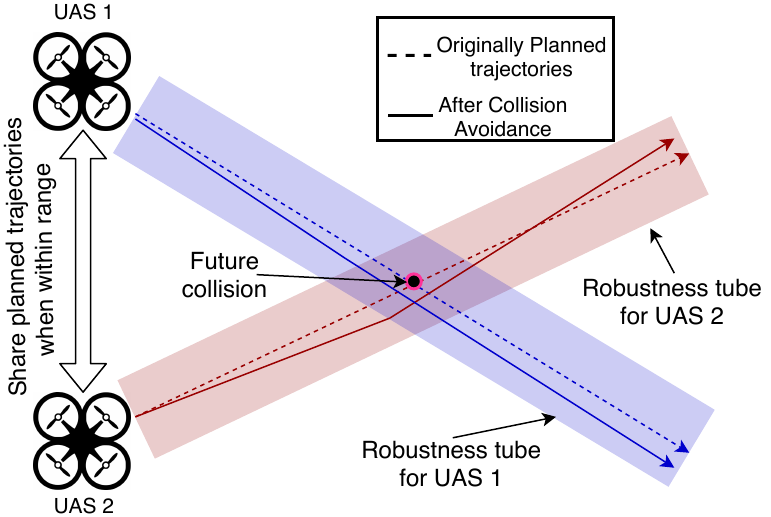}
\caption{\small Two UAS communicating their planned trajectories, and cooperatively maneuvering within their \textit{robustness tubes} to avoid a potential collision in the future.}
\vspace{-10pt}
\label{fig:CAdiagram}
\end{figure}



To overcome the limitations outlined above, we aim to solve the following problems: 

\begin{myprob}
\label{prob:traj_STL_highlevel}
\textbf{Independent trajectory planning} for an individual UAS (or fleets run by the same operator) to satisfy spatial, temporal and reactive mission objectives specified using Signal Temporal Logic (STL), independently of other UAS that could be in the airspace.
\end{myprob}

\begin{myprob}
\label{prob:CA_highlevel}
\textbf{Real-time pairwise predictive airborne collision avoidance} such that UAS mission requirements satisfied by the trajectories obtained by solving problem \ref{prob:traj_STL_highlevel} are not violated, e.g. UAS make it to their destinations in time while avoiding collisions with each other.
\end{myprob}




The airborne collision avoidance (Problem \ref{prob:CA_highlevel}) poses a larger challenge, and is the primary focus here. 
\begin{figure*}[t]
\includegraphics[width=0.99\textwidth]{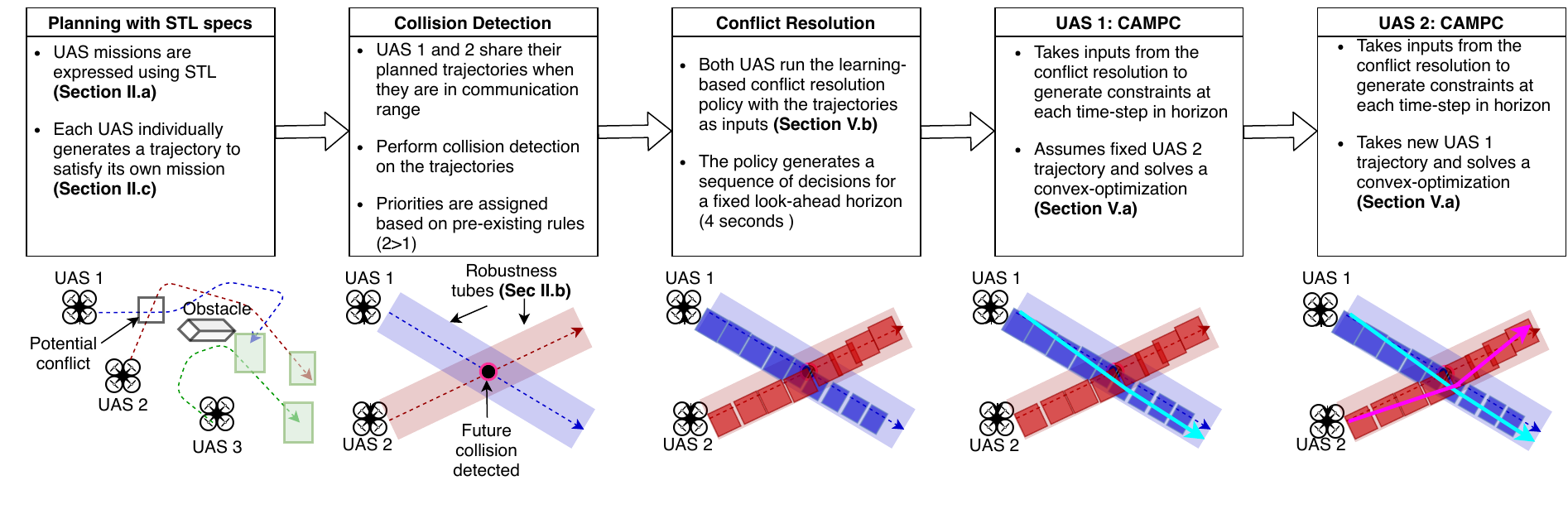}
\vspace{-10pt}
\caption{\small Step-wise explanation and visualization of the framework. Each UAS generates its own trajectories to satisfy a mission expressed as a Signal Temporal Logic (STL) specification, e.g. regions in green are regions of interest for the UAS to visit, and the obstacle corresponds to infrastructure that all the UAS must avoid. When executing these trajectories, UAS communicate their trajectories to others in range to detect any collisions that may happen in the near future. If a collision is detected, the two UAS execute a conflict resolution scheme that generates a set of additional constraints that the UAS must satisfy in order to avoid the collision. A co-operative CA-MPC controls the UAS in order to best satisfy these constraints while ensuring each UAS's STL specification is still satisfied. This results in new trajectories (in solid blue and pink) that will avoid the conflict and still stay within the pre-defined robustness tubes.}
\vspace{-15pt}
\label{fig:concept}
\end{figure*}

\input{chapters/contributions}

\input{chapters/overview}

\input{chapters/relatedWork}

%% file: chapters/contributions.tex
\subsection{Contributions of this work}

Our main contribution is \textit{Learning-to-Fly} (L2F)\footnote{Videos of the simulations and demonstrations in this paper can be viewed at \url{https://tinyurl.com/vvvuukh}}, a scheme 
for real-time, on-the-fly collision avoidance between two UAS whose main features are:

\begin{enumerate}
\item \textit{Systematic composition of machine learning and control theory:} We combine learning-based decision-making, and linear programming-based control to solve the problem in a decentralized manner. Unlike many other ad-hoc Machine Learning-based solutions, we provide a sound theoretical justification for our approach in Theorem \ref{th:MILP_CAMPC_relation}. We also provide a sufficient condition for the scheme to work successfully (Theorem \ref{th:CAMPC_success}).

\item  \textit{A notion of priority among the UAS} can be encoded naturally in L2F, where the UAS with higher priority does not have to deviate from its originally planned trajectory until absolutely necessary.

\item \textit{Computationally lightweight enough for real-time implementation:} Experimental results show that L2F, with a computation time in milliseconds can be used in a real-time implementation at a high-rate ($10$ Hz). 

\item \textit{High performance:} In the best case, L2F successfully results in 2-UAS collision avoidance $100\%$ of the test cases, gracefully degrading to $90\%$ for the worst case. Comparisons with other methods also show the superior performance of L2F. 

\item \textit{Enabling fast, independent planning for UAS with temporal logic objectives}, as individual UAS, or fleets of UAS run by the same operator, can plan for themselves without considering other UAS in the airspace while calling upon L2F for on-the-fly collision avoidance. For a 4-UAS case study, we demonstrate a speed up of $3.5\times$ over the centralized planning method of \cite{pant2018fly}.

\item \textit{Proof-of-concept demonstration} on Crazyflie quad-rotor robots to show feasibility on real UAS.

\end{enumerate}

%% file: chapters/overview.tex
\subsection{Overview of approach and paper outline}
\label{sec:overview}

In this paper, we aim to develop a framework for UAS traffic management (UTM) that solves problems \ref{prob:traj_STL_highlevel} and \ref{prob:CA_highlevel}.  Figure~\ref{fig:concept} depicts the proposed planning and control process and indicates the relevant sections in the paper.

\begin{enumerate}
\item 
\textit{Trajectory planning with Signal Temporal Logic (STL) specifications:} 
Each UAS, $j$, given the mission as an STL specification $\varphi_j$, generates a trajectory that robustly satisfies $\varphi_j$. The \emph{robustness value} $\rho_{\varphi_j}$, associated with this trajectory, corresponds to the maximum deviation from the planned trajectory such that the UAS $j$ still satisfies its mission $\varphi_j$.
\end{enumerate}

Two UAS within communication range share a look-ahead of planned trajectories and if a future collision is detected, new trajectories 
are needed that still satisfy their original mission specifications. 
For this, we develop our decentralized approach L2F, which consists of two stages:

\begin{enumerate}
  \setcounter{enumi}{1}
\item \textit{Collision detection} and \textit{Conflict resolution:} 
When a potential collision is detected, a supervised-learning based conflict resolution policy (CR-S), with pre-defined priority among the two UAS, generates a sequence of discrete decisions corresponding to maneuvers to avoid the collision. 

\item \textit{Distributed and co-operative Collision Avoidance MPC (CA-MPC):} The CA-MPC for each UAS takes as input the conflicting trajectories and the output of the conflict resolution policy, and controls the UAS to avoid collision. 
\end{enumerate}

In Section \ref{sec:experiments} we evaluate our framework for 2 UAS collision avoidance through extensive simulations and compare its performance to other approaches. 
Section \ref{sec:casestudy} demonstrates a particular UTM framework case study.  
Finally, in Section \ref{sec:future} we discuss potential future directions.


%% file: chapters/relatedWork.tex
\subsection{Related work}
\label{sec:relatedwork}
\subsubsection{UTM and Automatic Collision Avoidance approaches}
The UAS Traffic Management (UTM) problem has been studied in various contexts. 
In the NASA/FAA Concept of Operations document \cite{FAA2018UTM}, an airspace allocation scheme is outlined where individual UAS reserve airspace in the form of 4D polygons (space and time), and the polygons of different UAS are not allowed to overlap. Similarly, \cite{maxetal} presents a voxel-based airspace allocation approach. Our approach is less restrictive and allows overlaps in the 4D polygons, but performs maneuvers for collision avoidance when two UAS are on track to a collision (see Fig. \ref{fig:CAdiagram}). TCAS \cite{TCAS} and ACAS \cite{kochenderfer2012next} systems for collision avoidance in commercial aircrafts rely on transponders in the two aircrafts to communicate information for the collision avoidance modules. 
These generate recommendations for the pilots to follow and create vertical separation between aircrafts \cite{ACASX}. 
In the context of UAS, \cite{UTMTCL4} uses vehicle-to-vehicle communication and tree-search based planning to achieve collision avoidance. 
ACAS-Xu \cite{ACASXu}, an automatic collision avoidance scheme for UAS relies on a look-up table to provide high-level recommendations to two UAS that have potentially colliding trajectories. 
 It restricts desired maneuvers for CA to the vertical axis for cooperative traffic, and the horizontal axis for uncooperative traffic. 
While we consider only the cooperative case in this work, our method does not restrict CA maneuvers to any single axis of motion. Finally, in its current form, ACAS-Xu also does not take into account any higher-level mission objectives, unlike our approach. 
This excludes its application to low-level flights in urban settings, e.g. it can result in situations where ACAS-Xu recommends an action that avoids a nearby UAS but results in the primary UAS going close to a static obstacle. 
Our method avoids this as CA maneuvers are restricted to keeping UAS inside \textit{robustness tubes} (see Fig. \ref{fig:concept}) such that mission requirements are not violated. 
For this reason, ACAS-Xu is currently only being explored for large, high-flying UAS \cite{ACASXu} and is not directly applicable to the problem we study here.

\subsubsection{Multi-agent planning with temporal logic objectives}
Many approaches exist for the problem of planning for multiple robotic agents with temporal logic specifications. 
Most rely on abstract grid-based representations of the workspace \cite{SahaRSJ14, DeCastro17}, or abstract dynamics of the agents \cite{Drona,AksarayCDC16}.
\cite{MaICUAS16} combines a discrete planner with a continuous trajectory generator. 
Some methods \cite{4459804, 1582935, 1641832}   work for subsets of Linear Temporal Logic (LTL) that do not allow for explicit timing bounds on the mission requirements.
While \cite{SahaRSJ14} uses a subset of LTL, safe-LTL$_f$ that allows them to express reach-avoid specifications with explicit timing constraints. However, in addition to a discretization of the workspace, they also restrict motion to a simple, discrete set of motion primitives. 
The predictive control method of \cite{Raman14_MPCSTL} allows for using the expressiveness of the complete grammar STL for mission specifications. 
It handles a continuous workspace and linear dynamics of robots, however its reliance on mixed-integer encoding (similar to \cite{Saha_acc16,KaramanF11_LTLrouting}) for the STL specification limit use in planning/control for multiple agents in 3D workspaces as seen in \cite{pant2017smooth}. 
The approach of \cite{pant2018fly} instead relies on optimizing a smooth (non-convex) function for generating trajectories for fleets of multi-rotor UAS with STL specifications. 
In our framework, we use the planning method of \cite{pant2018fly}, but we let each UAS plan independently of each other. 
We ensure the safe operation of all UAS in the airspace through the use of our predictive collision avoidance scheme.




%% file: chapters/preliminaries.tex
\section{Preliminaries}
\label{sec:preliminaries}

We use Signal Temporal Logic (STL) to specify the mission objectives that the UAS need to satisfy (Problem \ref{prob:traj_STL_highlevel}). This section provides a brief introduction to STL and the trajectory generation approach.

\input{chapters/STL_short}

\input{chapters/stl_rob_short}
\input{chapters/problemStatement_planning}

%% file: chapters/STL_short.tex
\subsection{Introduction to Signal Temporal Logic}
\label{sec:STL_intro}
Signal Temporal Logic (STL) \cite{MalerN2004STL} is a behavioral specification language that can be used to encode requirements on signals. 
The grammar of STL~\cite{Raman14_MPCSTL} allows for capturing a rich set of 
behavioral requirements using temporal operators, such as \textit{Always} ($\always$) and \textit{Eventually} ($\eventually$), 
 as well as logical operators like \textit{And} ($\land$), \textit{Or} ($\lor$), and \textit{negation} ($\neg$). 
With these operators, an STL specification $\varphi$ is defined over a signal, e.g. over the trajectories of quad-rotor robots, and evaluates to either \textit{True} or \textit{False}. 
The following example demonstrates STL to capture operational requirements for two UAS:

\begin{exmp}
\label{ex:reach_avoid_exmp}
\textit{(A two UAS timed reach-avoid problem)} 
Two quad-rotor UAS are tasked with a mission with spatial and temporal requirements in the workspace shown in Fig. \ref{fig:dronetube}:

\begin{enumerate}

\item The two UAS have to reach a $\text{Goal}$ set (shown in green), 
or a region of interest, within a time of $6$ seconds after starting. 
UAS $j$ (where $j\in \{1,2\}$), with position denoted by $p_j$, has to satisfy: $\varphi_{reach, j} = 
\eventually_{[0,6]} (p_j \in \text{Goal})$.
The \textit{Eventually} operator over the time interval $[0,6]$ requires UAS $j$ to be inside the set $\text{Goal}$ at some point within $6$ seconds. 

\item In addition, the two UAS also have an $\text{Unsafe}$ (in red) set to avoid, e.g. a no-fly zone. For each UAS $j$, this is encoded with \textit{Always} and \textit{Negation} operators:

$\varphi_{\text{avoid},j} = \always_{[0,6]} \neg (p_j \in 
\text{Unsafe})$

\item Finally, the two UAS should also be separated by at least $\delta$ meters along every axis of motion:

$\varphi_{\text{separation}} = \always_{[0,6]} ||p_1 - p_2||_{\infty} 
\geq \delta$

\end{enumerate}

The 2-UAS timed reach-avoid specification is thus:
\begin{equation}
\label{eq:timed_RA}
\varphi_{\text{reach-avoid}} = \land_{j=1}^2 ( \varphi_{\text{reach},j} \land 
\varphi_{\text{avoid},j}) \land \varphi_\text{separation}
\end{equation}
\end{exmp}

In order to satisfy $\varphi$, a planning method generates trajectories $\mathbf{p}_1$ and $\mathbf{p}_2$ of a duration at least $hrz(\varphi)= 6$s, where $hrz(\varphi)$ is the time \textit{horizon} of $\varphi$. 
If the trajectories satisfy the specification, i.e. $(\mathbf{p}_1,\, \mathbf{p}_2) \models \varphi$, then the specification $\varphi$ evaluates to \textit{True}, otherwise it is \textit{False}. 
In general, an upper bound for the time horizon can be computed as shown in \cite{Raman14_MPCSTL}. 
In this work, we consider specifications such that the horizon is bounded. More details on STL can be found in \cite{MalerN2004STL} or \cite{Raman14_MPCSTL}. 
In this paper, we consider discrete-time STL semantics which are defined over discrete-time trajectories.


%
%
%
%

%% file: chapters/stl_rob_short.tex
\subsection{Robustness of STL specifications}
\label{sec:stl_rob_short}

For a time domain $\TDom = [0, T]$ with sampling time $dt$,
the signal space $\SigSpace$ is the set of all signals $\sstraj: \TDom \rightarrow X$. The \textit{Robustness} value \cite{FainekosP09tcs} $\rho_\formula$ of an STL formula $\formula$, with respect to the signal $\mathbf{x}$ that it is defined over, is a real-valued function of $\mathbf{x}$ that has the important following property:

\begin{theorem} \cite{FainekosP09tcs}
	\label{thm:rob objective new}
	For any $\sstraj \in \SigSpace$ and STL formula $\formula$, 
	if $\robf(\sstraj) <0$ then $\sstraj$ violates $\formula$, and if $\robf(\sstraj) > 0$ then $\sstraj$ satisfies $\formula$. 
	The case $\robf(\sstraj) =0$ is inconclusive.
\end{theorem} 
Intuitively, the degree of satisfaction or violation of a specification is indicated by the robustness value.
For simplicity, the  distances are defined in the inf-norm sense. 
This, combined with Theorem \ref{thm:rob objective new} gives us the following result:
\begin{corollary}
\label{cor:rob_tube}
Given a discrete-time trajectory $\sstraj$ such that $\sstraj \models \formula$ with robustness value $\rho>0$, then any trajectory $\mathbf{x}'$ that is within $\rho$ of $\sstraj$ at each time step, i.e. $||x_t-x_t'||_\infty < \rho \, \forall t \in \TDom$, is such that $\mathbf{x}' \models \formula$ (also satisfies $\formula$).
\end{corollary}


%% file: chapters/problemStatement_planning.tex
\subsection{UAS planning with STL specifications}
\label{sec:problem_planning}

Fly-by-logic \cite{pant2018fly} generates trajectories by centrally planning for fleets of UAS with STL specifications, e.g. the specification $\varphi_{\text{reach-avoid}}$ of example \ref{ex:reach_avoid_exmp}. 
It maximizes a smooth approximation $\srob_\formula$ of the robustness function \cite{pant2017smooth} by picking waypoints (connected via jerk-minimzing splines \cite{MuellerTRO15}) for all UAS through a centralized, non-convex optimization.

While successful in planning for multiple multi-rotor UAS, performance degrades as the number of UAS being planned for increases. The non-convex optimization involving the variables of all the UAS becomes harder as the number of variables increase \cite{pant2018fly} in particular because for $J$ UAS, $J \choose 2$ terms for pair-wise separation between the UAS are needed. 
Taking this into account, we use the underlying optimization of  \cite{pant2018fly} to generate trajectories
, but ignore the mutual separation requirement, allowing each UAS to independently (and in parallel) solve for their own STL specification. For the timed reach-avoid specification \eqref{eq:timed_RA} in example \ref{ex:reach_avoid_exmp}, this is equivalent to each UAS generating its own trajectory to satisfy $\varphi_j = \varphi_{reach, j} \land \varphi_{avoid, j}$, independently of the other UAS. Associated with these trajectories, $\sstraj_j$ is a robustness values $\rho_{\varphi_j}$. 
Ignoring the collision avoidance requirement ($\varphi_\text{separation}$) in the planning stage allows for the specification of \eqref{eq:timed_RA} to be decoupled across UAS, but now requires online pairwise UAS collision avoidance if the planned trajectories are in conflict. This is covered in the following section.

\textbf{Note:} 
In the following sections, $\mathbf{x}$ will refer to a full-state (discrete-time, finite duration) trajectory for a UAS. We will also use $\mathbf{p}$ to refer to the position components in that trajectory, the position trajectory. $x_k$ (and $p_k$) refer to the components of the trajectory at time step $k$.


%% file: chapters/problemStatement.tex
\section{Problem formulation: Collision Avoidance} \label{sec:CA}


While flying their planned trajectories (from the previous section), two UAS that are within communication range share a look-ahead of their trajectories and check for a potential collision at any time step $k$ in this look-ahead horizon of $N$ time steps.
We assume the UAS can communicate with each other in a manner that allows for enough advance notice for avoiding collisions, e.g. using 5G technology. 
The details of this are beyond the scope of this paper.


\begin{mydef}
	\label{def:msep}
	\textbf{2-UAS Conflict}: Two UAS, with discrete-time positions $\mathbf{p}_1$ and $\mathbf{p}_2$ are said to be in \textit{conflict} at time step $k$ if $||p_{1,k}-p_{2,k}||_\infty < \delta$, where $\delta$ is a predefined minimum separation distance\footnote{A more general polyhedral constraint of the form $H(p_{1,k}-p_{2,k})< g$ can be used for defining the conflict without loss of generality.}. Here, $p_{j,k}$ represents the position of UAS $j$ at time step $k$. 
\end{mydef}

\begin{figure}[tb]
	\begin{center}
	\includegraphics[width=0.35\textwidth,trim={6cm 0cm 7.5cm 3cm},clip]{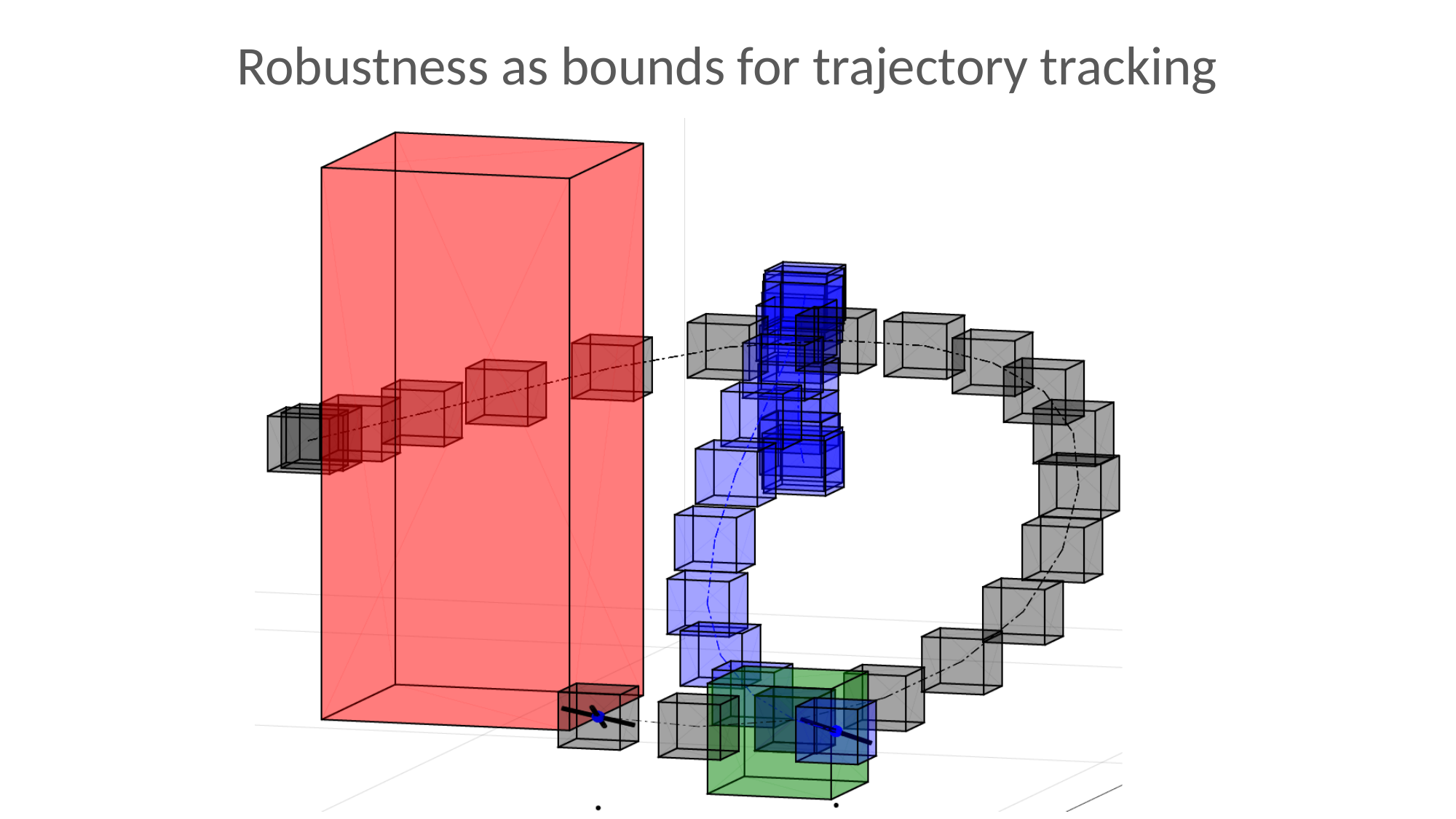}
	\end{center}
	\vspace{-10pt}
	\caption{\small Discrete time trajectories of two UAS, and their associated robustness tubes (see def. \ref{def:robustnesstube}) in gray and purple. The trajectories satisfy a reach-avoid specification, see example \ref{ex:reach_avoid_exmp}. $\text{Unsafe}$ set is in red and the $\text{Goal}$ set is in green.}
	\label{fig:dronetube}
	\vspace{-10pt}
\end{figure}

\begin{mydef}
	\label{def:robustnesstube}
	\textbf{Robustness tube}: Given an STL formula $\varphi$ and a discrete-time position trajectory $\mathbf{p}_j$ that satisfies $\varphi$ (with associated robustness $\rho$), the (discrete) \textit{robustness tube} around $\mathbf{p}_j$ is given by $\mathbf{P}_j = \mathbf{p}_j\oplus \mathbb{B}_{\rho_{\varphi}}$. We say the \textit{radius} of this tube is $\rho$ (in the inf-norm sense). Here $\mathbb{B}_\rho$ is a 3D cube with sides $2\rho$ and $\oplus$ is the Minkowski sum operation.
\end{mydef}
See an example of the robustness tube in Figure~\ref{fig:dronetube}. 

%

\textbf{Note}: As long as a UAS stays within its robustness tube, it will satisfy the STL specification $\varphi$ for the which the trajectory was generated for (see Corollary \ref{cor:rob_tube}). 

The following assumption now relates the minimum allowable radius $\rho$ of the robustness tube to the minimum allowable separation $\delta$ between two UAS.

\begin{myass}
	\label{assumption1}
	For each of the two UAS in conflict, the radius of the robustness tube is greater than $\delta/2$, i.e. $\min (\rho_1,\rho_2) \geq \delta/2$ where $\rho_1$ and $\rho_2$ are the robustness of UAS 1 and 2, respectively.
\end{myass}
This assumption defines the case where the radius of the robustness tube is just wide enough to have two UAS placed along opposing edges (of a cube at the same time step) and still achieve the minimum separation between them. We assume that all the trajectories generated by the independent planning have sufficient robustness to satisfy this assumption (see Sec. \ref{sec:problem_planning}).
Now we define the problem of collision avoidance with satisfaction of STL specifications:
\begin{myprob}
	\label{prob:deconfliction}
	Given two planned $N$-step UAS trajectories $\mathbf{p}_1$ and $\mathbf{p}_2$ that have a conflict, the collision avoidance problem is to find a new sequence of positions $\mathbf{p}_1'$ and $\mathbf{p}_2'$ that meet the following conditions:
	\begin{subequations}
		\begin{align}
		||p_{1,k}'-p_{2,k}'|| &\geq \delta \, \forall k = 0,\dotsc,N \label{eq:msep}\\
		p_{j,k}' &\in \mathbf{P}_j \, \forall j=1,2, \, \forall k = 0,\dotsc,N \label{eq:intube}
		\end{align}
	\end{subequations}
\end{myprob}
This implies that we need a new trajectory for each UAS such that they achieve minimum separation distance and also stay within the robustness tube around their originally planned trajectories (see Corollary \ref{cor:rob_tube}). 




\subsection{Convex constraints for collision avoidance}

Let $z_k$ be the difference in UAS positions at time step $k$. For two UAS not to be in conflict, we need 
\begin{equation}
\label{eq:noconf}
z_k = p_{1,k} - p_{2,k}  \not \in \mathbb{B}_\delta \, \forall k
\end{equation}
This is a non-convex constraint. For a computationally tractable controller formulation which solves problem \ref{prob:deconfliction}, we define convex constraints that when satisfied imply eq. \eqref{eq:noconf}. 

The $3$D cube $\mathbb{B}_\delta$ can be defined by a set of linear inequality constraints of the form $\widetilde{H}^i z \leq \widetilde{g}^i \, \forall i=1,\dotsc,6$.
Eq.~\eqref{eq:noconf} is satisfied when $\exists i \, | \widetilde{H}^i z > \widetilde{g}_i$. Let $H = -\widetilde{H}$ and $g = -\widetilde{g}$, then for any $i \in \{1,\dotsc,6\}$, 
\begin{equation}
\label{eq:pickaside}
H^i(p_{1,k}-p_{2,k}) < {g}^i \Rightarrow (p_{1,k}-p_{2,k}) \not \in \mathbb{B}_\delta 
\end{equation}

Intuitively, picking one $i$ at time step $k$ results in a configuration (in position space) where the two UAS are separated in one of two ways along one of three axis of motion\footnote{Two ways along one of three axis defines $6$ options, $i\in\{1,\ldots,6\}$.}, e.g. at a time step $k$ if we select $i|H^i=\begin{bmatrix}0& 0& 1\end{bmatrix},\, g^i = -\delta$, it implies than UAS 2 flies over UAS 1 by $\delta$ m, and so on.


\section{Centralized solution: MILP formulation}
\label{sec:subsec_milp}

Let the dynamics of either UAS\footnote{For simplicity we assume both 
	UAS have identical dynamics associated with multi-rotor robots, 
	however our approach would work otherwise.} be of the form $x_{k+1} = Ax_k + Bu_k$. The states 
$x_k \in \mathbb{R}^6$ here are the positions and velocities in the 
3D space, or $x_k = [p_k,\,v_k]^T$ (here 
$p$ and $v$ are the positions and velocities in the 3D space). The 
inputs $u_k \in \mathbb{R}^3$ are the thrust, roll and pitch of the 
UAS. The matrices $A$ and $B$ can be obtained through linearization 
of the UAS dynamics around hover and discretization in time \cite{PantAMNDM15_Anytime}. 
Let $C$ be the observation matrix such that $p_k=Cx_k$. 

For $N$ steps into the future with a conflict, solving the following receding horizon MILP over the variables of the two UAS would result in new trajectories $\mathbf{p}_1', \, \mathbf{p}_2'$ that satisfy the minimum separation requirement \eqref{eq:noconf}. 
Let $\mathbf{x}_j \in \mathbb{R}^{6(N+1)}$ be the pre-planned full state trajectories, $\mathbf{x}_j' \in \mathbb{R}^{6(N+1)}$ the new full state trajectories and $\mathbf{u}_j' \in \mathbb{R}^{3N}$ the new controls to be computed for the two UAS ($j=1,\,2$). 
Let $\mathbf{b} \in \{0,1\}^{6(N+1)}$ be binary decision variables, and $M$ is a large positive number, then the MILP problem is defined as:
\begin{equation}
\label{eq:CentralMILP}
\resizebox{.442\textwidth}{!}{%
	$
	\begin{aligned}
	\min_{\mathbf{u}_1', \, \mathbf{u}_2', \, \mathbf{b} |\mathbf{x}_1,\, \mathbf{x}_2} &L(\mathbf{x}_1', \, \mathbf{u}_1', \, \mathbf{x}_2', \, \mathbf{u}_2') \\
	x_{j,0}' &= x_{j,0} \, \forall j \in \{1,2\} \\
	x_{j,k+1}' &= Ax_{j,k}' + Bu_{j,k}' \, \forall k \in \{0,\dotsc,N-1\} , \, \forall j \in \{1,2\}\\
	Cx'_{j,k} &\in P_{j,k} \, \forall k \in \{0,\dotsc,N\} , \, \forall j \in \{1,2\}\\
	H^{i}C(x_{1,k}'\!-\!x_{2,k}') &\leq {g}_{i} \!+\! M(1\!-b_{i,k}\!) \, \forall k \in \{0,\dotsc\!,N\}, \forall i \in \{1,\dotsc,\!6\} \\
	\sum_{i=1}^6 b^i_{k} &\geq 1 \, \forall k \in \{0,\dotsc,N\} \\
	u_{j,k}' &\in U \, \forall k \in \{0,\dotsc,N\} , \, \forall j\in \{1,2\}\\
	x_{j,k}' &\in X \, \forall k \in \{0,\dotsc,N+1\}, \forall j\in \{1,2\}
	\end{aligned}$%
}
\end{equation}
Here $b^i_k$ encodes action $i=1,\dotsc,6$ taken for avoiding a collision at time step $k$ which corresponds to a particular side of the cube $\mathbb{B}_\delta$.
A solution (when it exists) to this MILP results in new 
trajectories that avoid collisions and stay within their respective robustness tubes of the original trajectories. 
However, this method relies on solving for a pair of UAS in a centralized manner. 
Also, it introduces $6$ times as many variables (and constraints) as the time horizon of the optimization, which could make the MILP computationally intractable for a real-time implementation. 
Therefore, we develop a decentralized approach in the following sections.


\section{Decentralized solution: Learning-to-Fly}
\label{sec:CA_MPC}

The distributed and co-operative collision avoidance MPC scheme of Section \ref{sec:ca_mpc} with the conflict resolution algorithm described in Section \ref{sec:learning_supervised} form the online collision avoidance scheme, \textit{Learning-to-Fly} (L2F), our main contribution.

We assume that the two UAS can communicate their pre-planned $N$-step trajectories $\mathbf{p}_1,\, \mathbf{p}_2$ to each other (refer to Sec. \ref{sec:problem_planning}).
Instead of solving the centralized MILP, we want to solve problem \ref{prob:deconfliction} by following these steps:

\begin{enumerate}
\item \textbf{Conflict resolution:} UAS 1 and 2 make a \textit{sequence of decisions}, $\mathbf{d}=(d_0,\ldots,d_N)$ to avoid collision. Each $d_k\in\{1,\ldots\,6\}$ 
represents a particular choice of $H$ and $g$ at time step $k$, see eq.~\ref{eq:pickaside}.
Section \ref{sec:learning_supervised} describes our proposed learning-based method for conflict resolution.
\item \textbf{UAS 1 CA-MPC:} UAS 1 takes the conflict resolution sequence $\mathbf{d}$ from step 1 and solves a convex optimization to try to deconflict while assuming UAS 2 maintains its original trajectory. After the optimization the new trajectory for UAS 1 is sent to UAS 2.
\item \textbf{UAS 2 CA-MPC:} (If needed) UAS 2 takes the same conflict resolution sequence $\mathbf{d}$ from step 1 and solves a convex optimization to try to avoid UAS 1's new trajectory. 
Section~\ref{sec:ca_mpc} provides more details on CA-MPC steps 2 and 3.
\end{enumerate}


The overall algorithm is shown in Alg. \ref{alg:l2f}. 
The visualization of the above steps is presented in Fig.~\ref{fig:concept}.
Such decentralized approach differs from the centralized MILP approach, where both the binary decision variables and continuous control variables for each UAS are decided concurrently. 

\subsection{Distributed and co-operative collision avoidance MPC}
\label{sec:ca_mpc}
Each UAS $j\in\{1,2\}$ solves the following Collision Avoidance MPC optimization\footnote{Enforcing the separation constraint at each time step can lead to a restrictive formulation, especially in cases where the two UAS are only briefly close to each other. This does however give us an optimization with a structure that does not change over time, and can avoid collisions in cases where the UAS could run across each other more than once in quick succession (e.g. \url{https://tinyurl.com/uex7722}), which is something ACAS-Xu was not designed for.}:


\textbf{$\text{CA-MPC}_j(\mathbf{x}_j,\, \mathbf{x}_{avoid},\, P_j,\ \mathbf{d},\, prty_j)$}:
\begin{equation}
\label{eq:campc}
\resizebox{.44\textwidth}{!}{
	$
	\begin{aligned}
	& \min_{\mathbf{u}_j',\mathbf{\lambda}_j|\mathbf{x}_j,\mathbf{x}_{avoid}} \sum_k \lambda_{j,k} \\
	x_{j,0}' &= x_{j,0} \\
	x_{j,k+1}' &= Ax_{j,k}' + Bu_{j,k}' \, \forall k=\{0,\dotsc,N-1\} \\
	Cx_{j,k}' &\in P_{j,k} \, \forall k=\{0,\dotsc,N\} \\
	prty_j \!\cdot\! H^{d_k}C(x_{avoid,k}\!-x_{j,k}') &\leq g^{d_k} \!+\! \lambda_{j,k}\, \forall k=\{0,\dotsc,N\} \\
	\lambda_{j,k} &\geq 0\, \forall k=\{0,\dotsc,N\} \\
	u_{j,k}' &\in U\, \forall k=\{0,\dotsc,N \} \\
	x_{j,k}' &\in X \, \forall k \in \{0,\dotsc,N+1\}
	\end{aligned}
	$
}
\end{equation}
where, $\mathbf{x}_j$ is the pre-planned trajectory of UAS $j$, $\mathbf{x}_{avoid}$ is the pre-planned trajectory  from which UAS $j$ must attain a minimum separation, $prty_j \in \{-1, +1\}$ is the priority of UAS $j$ w.r.t the other UAS in conflict.
The decision sequence $\mathbf{d}$ is represented as $H^{d_k},\, g^{d_k}$. 
This MPC optimization tries to find a new trajectory $\mathbf{x}_j'$ for the UAS $j$ that minimizes the slack variables $\lambda_{j,k}$ that correspond to violations in the minimum separation constraint $\eqref{eq:pickaside}$ w.r.t the pre-planned trajectory $\mathbf{x}_{avoid}$ of the UAS in conflict. 
The constraints in \eqref{eq:campc} ensure that UAS $j$ respects its dynamics, input constraints, and state constraints to stay inside the robustness tube. 
An objective of $0$ implies that UAS $j$'s new trajectory satisfies the minimum separation between the two UAS, see eq.~\eqref{eq:pickaside}.

\textbf{CA-MPC optimization for UAS 1:}
UAS 1, with lower priority, $prty_1 = -1$, first attempts to resolve the conflict for the given sequence of decisions $\mathbf{d}$.
An objective of $0$ implies that UAS 1 alone can satisfy the minimum separation between the two UAS. 
Otherwise, UAS 1 alone could not create separation and UAS 2 now needs to maneuver as well.

\textbf{CA-MPC optimization for UAS 2:}
If UAS 1 is unsuccessful at collision avoidance, UAS 1 communicates its current revised trajectory $\mathbf{x}_1'$ to UAS 2, with $prty_2 = +1$.
UAS 2 then creates a new trajectory $\mathbf{x}_2'$ 
(w.r.t the same decision sequence $\mathbf{d}$).

Alg. \ref{alg:l2f} is designed to be computationally lighter than the 
MILP approach (see Section~\ref{sec:subsec_milp}), but unlike the MILP it is not complete. 
In Section \ref{sec:experiments}, through extensive simulations we show that the L2F approach demonstrates a significant improvement in runtime while  maintaining comparable performance in terms of separation.

\begin{algorithm}
	\KwData{Pre-planned trajectories, robustness tubes}
	\KwResult{Sequence of control signals $\mathbf{u}_1'$, $\mathbf{u}_2'$ for the two UAS}
	Get $\mathbf{d}$ from conflict resolution 
	
	UAS 1 solves CA-MPC optimization \eqref{eq:campc}:\\
	$(\mathbf{x}_1', \mathbf{u}_1', \mathbf{\lambda}_1)=\textbf{CA-MPC}_1(\mathbf{x}_1,\mathbf{x}_2, P_1, \mathbf{d}, -1)$ 
	
	\eIf{$\sum_k \lambda_{1,k} = 0$}
	{\textbf{Done}: UAS 1 alone has created separation; Set $\mathbf{u}_2'=\mathbf{u}_2$
	}
	{ UAS 1 transmits solution to UAS 2
		
		UAS 2 solves CA-MPC optimization \eqref{eq:campc}: \\
		$(\mathbf{x}_2', \mathbf{u}_2', \mathbf{\lambda}_2)=\textbf{CA-MPC}_2(\mathbf{x}_2,\mathbf{x}_1', P_2, d, +1)$
		
		\eIf{$\sum_k \lambda_{2,k} = 0$}{\textbf{Done:} UAS 2 has created 
			separation }
		{\eIf{$||p_{1,k}'-p_{2,k}'|| \geq \delta \, \forall k = 0,\dotsc,N$}
			{\textbf{Done}: UAS 1 and UAS 2 created separation}
			{\textbf{Not done}: UAS still violate eq. \ref{eq:msep}}}
	}
	
	Apply control signals $\mathbf{u}_1'$, $\mathbf{u}_2'$ if \textbf{Done}; else \textbf{Fail}.
	\caption{Learning-to-Fly: Decentralized and cooperative collision avoidance for two UAS. Also see fig. \ref{fig:concept}.}
	\label{alg:l2f}
	
\end{algorithm}
The solution of CA-MPC can be defined as follows:
\begin{definition}[Zero-slack solution]
	\label{def:zero_slack}
	The solution of the CA-MPC optimization \eqref{eq:campc}, is called the \textit{zero-slack solution} if for a given decision sequence $\mathbf{d}$ either
	
	1) there exists an optimal solution of \eqref{eq:campc} such that  $\sum_k\lambda_{1,k}=0$ or
	
	2) problem \eqref{eq:campc} is feasible with $\sum_k\lambda_{1,k}>0$ and there exists an optimal solution of \eqref{eq:campc} such that  $\sum_k\lambda_{2,k}=0$.
\end{definition}

The two following theorems make important connections between feasible solutions for MILP and CA-MPC formulations.
They are the consequence of the construction of CA-MPC optimizations. We omit the proofs for brevity.

\begin{theorem}[Sufficient condition for CA]
	\label{th:CAMPC_success}
	Zero-slack solution of \eqref{eq:campc}
	implies that the resulting trajectories for two UAS are non-conflicting and within the robustness tubes of the initial trajectories\footnote{Theorem~\ref{th:CAMPC_success} formulates a conservative result as \eqref{eq:pickaside} is a convex under approximation of the originally non-convex collision avoidance constraint \eqref{eq:noconf}. Indeed, non-zero slack $\exists k| \lambda_{2,k}>0$ does not necessarily imply the violation of the mutual separation requirement \eqref{eq:msep}. The control signals $u_1',u_2'$ computed by alg. \ref{alg:l2f} can therefore in some instances still create separation between drones even when the conditions of Theorem \ref{th:CAMPC_success} are not satisfied.}.
\end{theorem}

\input{chapters/feasMILPCAMPC}

\input{chapters/learning_supervised}

%% file: chapters/feasMILPCAMPC.tex
\begin{theorem}[Existence of the zero-slack solution]
	\label{th:MILP_CAMPC_relation}
	Feasibility of the MILP problem~\eqref{eq:CentralMILP} implies the existence of the zero-slack solution of CA-MPC optimization \eqref{eq:campc}.
\end{theorem}

The Theorem~\ref{th:MILP_CAMPC_relation} states that the binary decision variables $b^i_k$ selected by the feasible solution of the MILP problem \eqref{eq:CentralMILP}, when used to select the constraints (defined by $H,\,g$) for the CA-MPC formulations for UAS 1 and 2, imply the existence of a zero-slack solution of \eqref{eq:campc}.

%


%% file: chapters/learning_supervised.tex
\subsection{Learning-based conflict resolution}
\label{sec:learning_supervised}

Motivated by Theorem~\ref{th:MILP_CAMPC_relation},
we propose to learn the conflict resolution policy 
from the MILP solutions.
To do so, we use a \textit{Long Short-Term Memory} (LSTM)~\cite{hochreiter1997long} recurrent neural network augmented with fully-connected layers.
LSTMs perform better than traditional recurrent neural networks on sequential prediction tasks~\cite{gers2002learning}.
 
 \begin{figure}[tb]
 	\begin{center}
 		\includegraphics[width=0.49\textwidth, trim={0cm 0.5cm 0cm 0cm}]{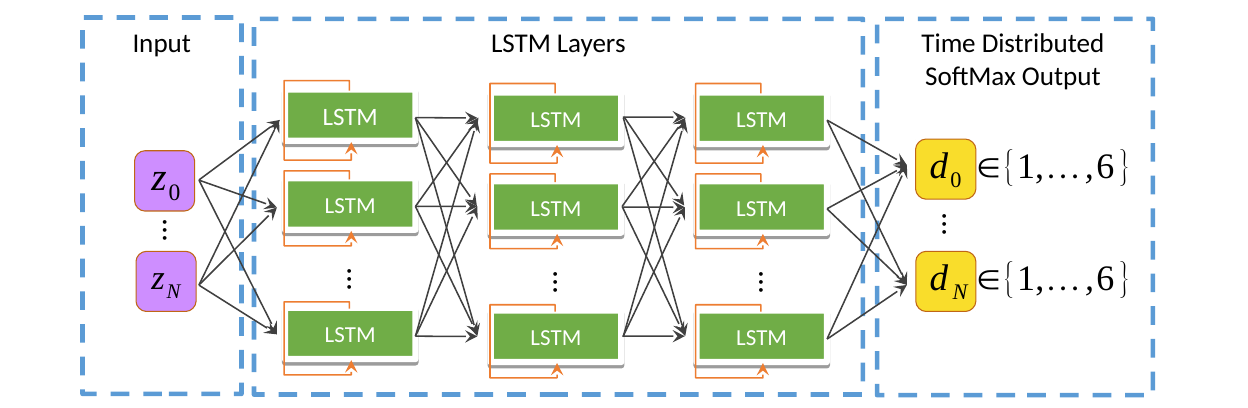}
 	\end{center}
 	{\caption{\small Proposed LSTM model architecture for CR-S. LSTM layers are shown unrolled over $N$ time steps. The inputs are 
 			$z_k$ which are the differences between the planned UAS positions, and the outputs are decisions $d_k$ for conflict resolution at each time $k$ in the horizon.}
 		\label{fig:lstm_arc}}
 	\vspace{-15pt}
 \end{figure} 

The network is trained to map a difference trajectory $\mathbf{z}=\mathbf{x}_1-\mathbf{x}_2$ (as in eq.~\eqref{eq:noconf}) to a decision sequence $\mathbf{d}$ that deconflicts pre-planned trajectories $\mathbf{x}_1$ and $\mathbf{x}_2$. For creating the training set, $\mathbf{d}$ is produced by solving the MILP problem~\ref{eq:CentralMILP}, i.e. obtaining 
a sequence of binary decision variables $\mathbf{b}\in\{0,1\}^{6(N+1)}$ and translating it into the decision sequence $\mathbf{d}\in\{1,\ldots,6\}^{N+1}$. 


The proposed architecture is presented in 
 Figure~\ref{fig:lstm_arc}.
The input layer is 
connected to the block of three stacked LSTM layers.
The output layer is a time distributed dense layer with a 
softmax activation function such that each value is a  
decision $d_k$, $k=\{0,\ldots,N\}$. 

%% file: chapters/experimental.tex
\section{Experimental evaluation}
\label{sec:experiments}

\input{chapters/implementation}

\input{chapters/experimental_results}

%% file: chapters/implementation.tex
\subsection{Experimental setup}

All the simulations were performed on a computer with an AMD Ryzen 7 2700 8-core processor and 16GB RAM, running Ubuntu 18.04. 
The MILP formulation was implemented in MATLAB using Yalmip \cite{lofberg2004yalmip} with MOSEK v8 as the solver. The learning-based approach was implemented in Python 3 with Tensorflow 1.14 and Keras API and Casadi with QPOASES as the solver. 
We implemented the CA-MPC using CVXGEN for a measurement of 
computation times and real-time implementation for experiments of actual hardware.



For the experiments, we set minimum separation to $\delta = 0.1$m. 
The learning-based CR scheme was trained for $\rho = 0.055$ which is close to the lower bound in assumption \ref{assumption1}.

We have generated the data set of 14K training and 10K test conflicting trajectories using the minimum-jerk trajectory generation algorithm from~\cite{pant2018fly}.
The time horizon was set to $T=4$s and $dt=0.1$s. 
The initial and final waypoints were sampled uniformly at random from two 3D cubes close to the fixed collision point, initial velocities were set to zero.


We have trained and ran experiments for various network 
configurations. 
For each model, the number of training epochs was set to 2K with a batch size of 2K. Each network was trained to minimize categorical cross-entropy loss using Adam optimizer with training rate of $0.001$.
The model with 3 LSTM layers with 128 neurons each has 
been chosen as the default learning-based CR model.

%% file: chapters/experimental_results.tex
\subsection{Results and comparison to other methods}
\label{sec:exp_results}

We analyzed three other methods alongside the proposed learning-based approach for L2F.
\begin{enumerate}
	\item A \textbf{random} decision approach which outputs a 
	sequence sampled from the discrete uniform distribution.
	\item A \textbf{greedy} approach that selects the discrete decisions for which the most distance between the two UAS available at each time step. 
	\item A centralized \textbf{MILP} solution that picks a decision corresponding to a binary decision variable in \eqref{eq:CentralMILP}.
\end{enumerate}

For the evaluation, we measured and compared the \textbf{separation rate} and the \textbf{computation time} over 10K test trajectories.
\textit{Separation rate} defines the fraction of given initially conflicting trajectories for which UAS managed to achieve minimum separation.

\begin{figure}[t!]
	\begin{center}
		\includegraphics[width=0.49\textwidth, trim={0 0.5cm 0 0.25cm}]{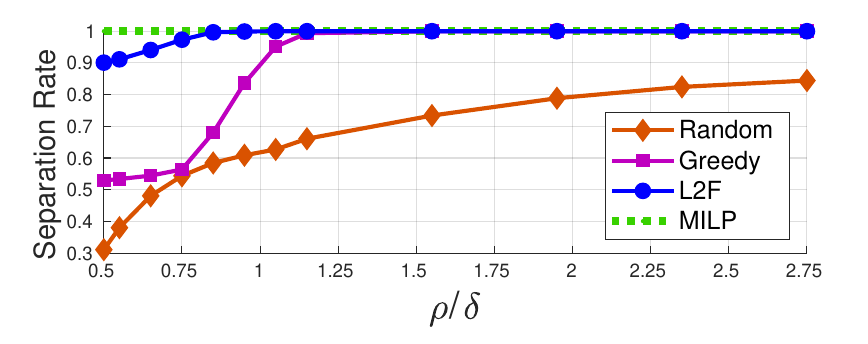}
	\end{center}
	\vspace{-5pt}
	{\footnotesize
		\caption{\small Model sensitivity analysis with respect to variations of fraction $\rho/\delta$, which connects the minimum allowable robustness tube radius $\rho$ to the minimum allowable separation between two UAS $\delta$, see Assumption~\ref{assumption1}. A higher $\rho/\delta$ implies there is more room within the robustness tubes to maneuver within for CA.}
		\label{fig:rho_rate_33}}
	\vspace{-10pt}
\end{figure} 

Figure~\ref{fig:rho_rate_33} shows the trade-off between performance in terms of separation rate and $\rho/\delta$ fraction, which defines the connection between the robustness tube $\rho$ and the minimum separation $\delta$. Higher $\rho/\delta$ implies wider robustness tubes for the UAS to maneuver within, which should make the CA task easier. In the case of $\rho/\delta=0.5$, where the robustness tubes are just wide enough to fit two UAS (see assumption~\ref{assumption1}), we see the L2F significantly outperforms the methods (excluding the MILP). As the ratio grows, the performance of all methods improve with L2F still outperforming the others, topping out to achieve a best case separation of $1$. The worst-case performance for L2F is $0.9$ which is again significantly better than the other approaches. 

Table \ref{tbl:success-time} 
shows the separation rates for three different $\rho/\delta$ values 
as well as the computation times for conflict resolution schemes plus the CA-MPC optimizations. In terms of separation rate, L2F outperformed the random and the greedy approaches. The centralized MILP outperformed the L2F, however, the computation time for the centralized approach was orders of magnitude higher than L2F. 
These shows the benefits of L2F compared to 
other approaches, especially when considering the success-computation time 
trade-off. 



\begin{table}[h]
	\vspace{-10pt}
	\renewcommand{\arraystretch}{1.3}
	\setlength{\tabcolsep}{2.5pt}
	\centering
	\begin{tabular}{|l||c|c|c||c|c|}
		\hline
		\multirow{2}{*}{\textbf{CA Scheme}} & \multicolumn{3}{c||}{\textbf{Separation Rate}} & \multicolumn{2}{c|}{\textbf{Computation time}} \\ \cline{2-6} 
		& \textbf{$\rho/\delta$ = 0.5}           & \textbf{$\rho/\delta$ = 0.95} & \textbf{$\rho/\delta$ = 1.15}     & \textbf{Mean}           & \textbf{Std}          \\ \hline\hline
		\textbf{Random}                     & 0.311                          & 0.609  &0.661           & 2.02ms                  & 0.17ms               \\ \hline
		\textbf{Greedy}                     & 0.529                         & 0.836  &0.994           & 3.82ms                  & 0.25ms               \\ \hline
		\textbf{L2F}                        & 0.901                         & 0.999   &1          & 9.36ms                  & 1.75ms               \\ \hline
		\textbf{MILP}                       & 1                             & 1   &1          & 68.5s                   & 87.3s                \\ \hline
	\end{tabular}
	\caption{{\small Separation rates and computation times (mean and standard deviation) comparison of different CA schemes. \textit{Separation rate} is the fraction of conflicting trajectories for which separation requirement
			\eqref{eq:msep} is satisfied after CA. 
	}}
	\label{tbl:success-time}
	\vspace{-5pt}
\end{table}

\begin{figure}[tb]
	\begin{center}
		\includegraphics[width=0.49\textwidth,trim={0 1cm 0 0.25cm}]{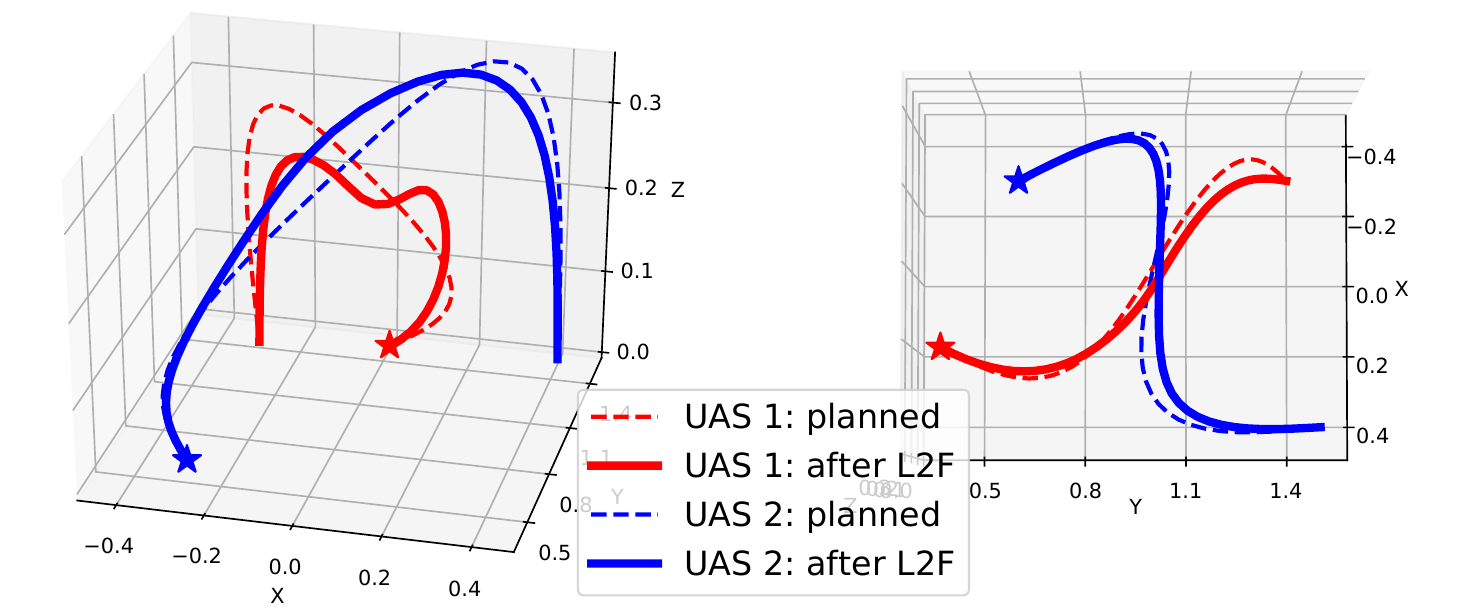}
	\end{center}
	\vspace{-2pt}
	{\footnotesize
		\caption{\small Trajectories for 2 UAS from different angles. The dashed (planned) trajectories have a collision at the halfway point. The solid ones, generated through L2F method, avoid the collision while remaining within the robustness tube of the original trajectories. Initial UAS positions marked as stars. Playback of the scenario is at \url{https://tinyurl.com/y8cm65ya}.}
		\label{fig:scen1_w_ca}\vspace{-10pt}}
\end{figure} 

Figure~\ref{fig:scen1_w_ca} shows an example of two UAS trajectories before and after collision avoidance through L2F method.
In addition, in order to evaluate the feasibility of the deconflicted trajectories, 
we have also ran experiments using two Crazyflie quad-rotor robots.   
Video recordings of the actual flights
and additional simulations can be found at {\footnotesize\url{https://tinyurl.com/yxttq7l5}}.

%% file: chapters/casestudy.tex
\section{Case study: Independent planning and L2F for a 4-UAS example}
\label{sec:casestudy}

\begin{figure}[h]
	\includegraphics[width=0.49\textwidth,trim={0 0cm 0.2cm 4.9cm},clip]{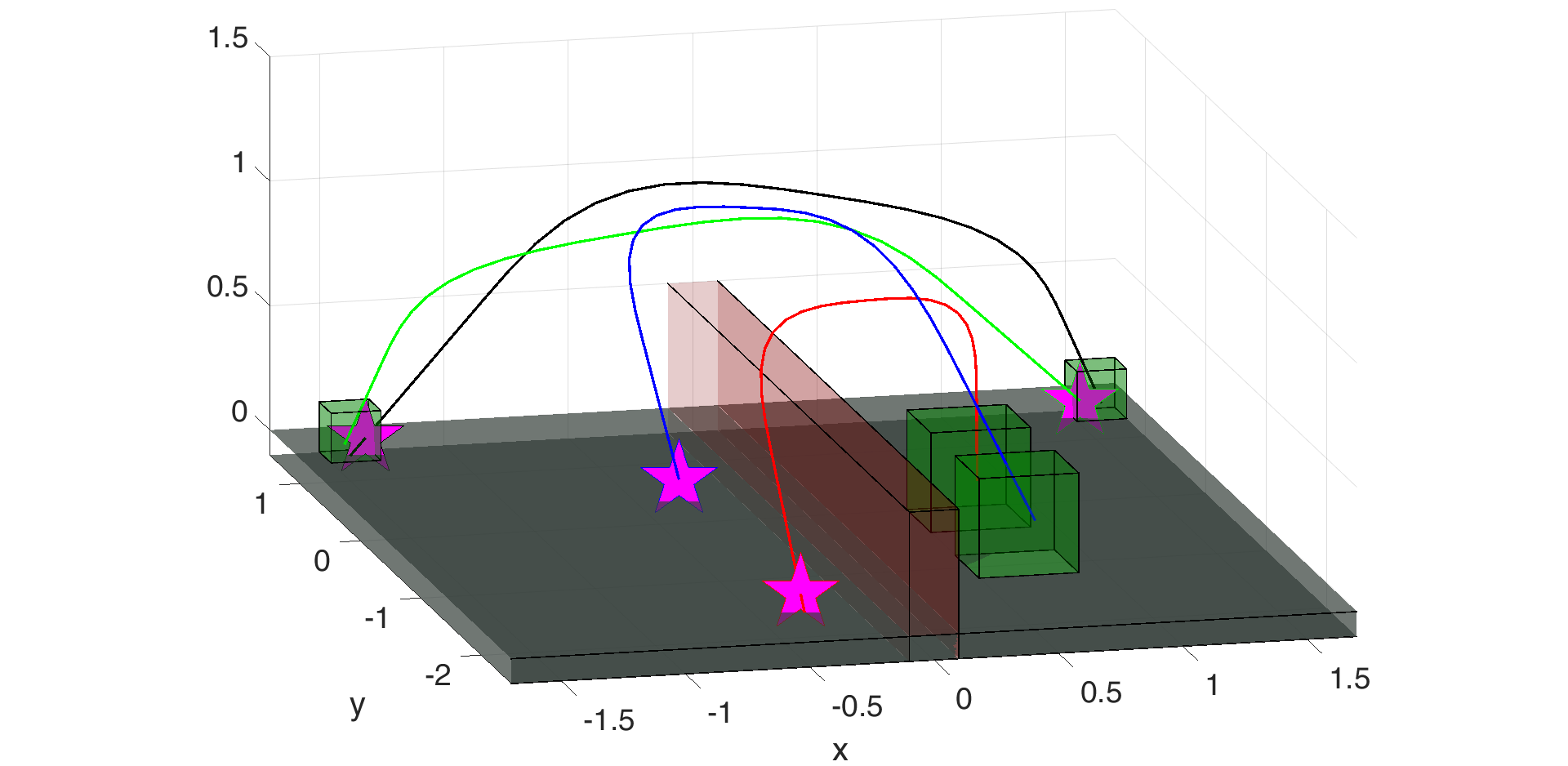}
	\vspace{-10pt}
	\caption{\small Workspace for the case study scenario. 
		Trajectories for 4 UAS (magenta stars) reaching their goal sets 
		(green boxes) within 4 seconds, while not crashing into the vertical 
		wall (in red). A pairwise separation requirement 
		of $0.1$m is enforced. Simulations are available at 
		\url{https://tinyurl.com/t8bwwqk}.}
	\label{fig:scenario}
	\vspace{-10pt}
\end{figure}

Figure~\ref{fig:scenario} depicts a UAS case-study with a 
reach-avoid mission. Scenario consists of four UAS which must 
reach desired goal states within 4 seconds while avoiding the wall 
obstacle and each other. Each UAS $j\in\{1,\ldots,4\}$ specification can be defined as:
\begin{equation}
\label{eq:scenario_spec_d}
\varphi_j =  
\eventually_{[0,4]} (p_j \in \text{Goal})
\ \wedge\ 
\always_{[0,4]} \neg (p_j \in \text{Wall})
\vspace{-2pt}
\end{equation}
A pairwise separations requirement of $0.1$ meters is enforced for 
all UAS, therefore, the overall mission specification is:
\vspace{-2pt}
\begin{equation}
\label{eq:case_mission}
\varphi_{\text{mission}} = \bigwedge_{j=1}^4 \varphi_j\ \wedge\ 
\bigwedge_{j\not=j'} \always_{[0,4]}||p_j-p_{j'}||\geq 0.1
\vspace{-3pt}
\end{equation}

%

First, we solved the planning problem for all four UAS in a centralized manner following 
approach from~\cite{pant2018fly} 
Next, we solved the planning problem for each UAS $j$ and its specification $\varphi_j$ independently, with calling L2F on-the-fly, after planning is complete. This way, independent planning with the online collision avoidance scheme guarantees the satisfaction of the overall mission specification \eqref{eq:case_mission}.

\textbf{Simulation results.}
We have simulated the scenario for 100 different initial conditions.
The average computation time to generate trajectories in a centralized manner was $0.35$ seconds. The average time per UAS when planning independently (and in parallel) was $0.1$ seconds. 
These results demonstrate a speed up of $3.5\times$ for the individual UAS planning versus centralized \cite{pant2018fly}. 




%

%% file: chapters/futurework.tex
\section{Conclusion}
\label{sec:future}

\textbf{Summary.}
We developed \textit{Learning-to-Fly} (L2F), a two-stage, on-the-fly and predictive collision avoidance approach that combines learning-based decision-making for conflict resolution with decentralized linear programming-based UAS control. 
Through extensive simulations and demonstrations on real quadrotor drones we show that L2F, with a run-time $<10$ms is computationally fast enough for real-time implementation. It is successful in resolving $100\%$ of collisions in most cases, with a graceful degradation to the worst-case performance of $90\%$ when there is little room for the UAS to maneuver. 
L2F also enables independent UAS planning, 
 speeding up the process compared to centrally planning for all the UAS in the airspace. A 4-UAS case study shows that the independent planning is $3.5\times$-faster. 

\textbf{Limitations and Future Work.}
While pairwise collision avoidance is sufficient when the airspace density is low, in the future we will extend the approach to cases where more than two UAS could be in conflict with each other. 
As L2F does not always succeed, we plan to investigate this further and use the failure cases as counterexamples to make the learning-based models better. In general, we expect L2F to be realized within a larger UTM system with additional contingencies (e.g. FAA Lost Link procedures \cite{pastakia_faa_2015}), including the possibility of online re-planning of missions when L2F cannot guarantee collision avoidance.

%% file: root.bbl
\begin{thebibliography}{10}
\providecommand{\url}[1]{#1}
\csname url@samestyle\endcsname
\providecommand{\newblock}{\relax}
\providecommand{\bibinfo}[2]{#2}
\providecommand{\BIBentrySTDinterwordspacing}{\spaceskip=0pt\relax}
\providecommand{\BIBentryALTinterwordstretchfactor}{4}
\providecommand{\BIBentryALTinterwordspacing}{\spaceskip=\fontdimen2\font plus
\BIBentryALTinterwordstretchfactor\fontdimen3\font minus
  \fontdimen4\font\relax}
\providecommand{\BIBforeignlanguage}[2]{{%
\expandafter\ifx\csname l@#1\endcsname\relax
\typeout{** WARNING: IEEEtran.bst: No hyphenation pattern has been}%
\typeout{** loaded for the language `#1'. Using the pattern for}%
\typeout{** the default language instead.}%
\else
\language=\csname l@#1\endcsname
\fi
#2}}
\providecommand{\BIBdecl}{\relax}
\BIBdecl

\bibitem{NAP25646}
{National Academies of Sciences, Engineering, and Medicine}, \emph{{Advanced
  Aerial Mobility: A National Blueprint}}.\hskip 1em plus 0.5em minus
  0.4em\relax The National Academies Press, 2020.

\bibitem{pant2018fly}
Y.~V. Pant, H.~Abbas, R.~A. Quaye, and R.~Mangharam, ``Fly-by-logic: control of
  multi-drone fleets with temporal logic objectives,'' in \emph{Proceedings of
  the 9th ACM/IEEE International Conference on Cyber-Physical Systems}.\hskip
  1em plus 0.5em minus 0.4em\relax IEEE Press, 2018, pp. 186--197.

\bibitem{SahaRSJ14}
I.~Saha, R.~Rattanachai, V.~Kumar, G.~J. Pappas, and S.~A. Seshia, ``Automated
  composition of motion primitives for multi-robot systems from safe ltl
  specifications,'' in \emph{IEEE/RSJ International Conference on Intelligent
  Robots and Systems}, 2014.

\bibitem{TCAS}
{Federal Aviation Administration}, ``{Introduction to TCAS II, V7.1},''
  \url{https://www.faa.gov/documentlibrary/media/advisory_circular/tcas%20ii%20v7.1%20intro%20booklet.pdf},
  accessed: 2019-10-19.

\bibitem{FAA2018UTM}
{Federal Aviation Administration (FAA)}, ``{Concept of Operations: Unmanned
  Aircraft System (UAS) Traffic Management (UTM)},''
  \url{{https://utm.arc.nasa.gov/docs/2018-UTM-ConOps-v1.0.pdf}}, 2018.

\bibitem{maxetal}
M.~Z. {Li}, W.~R. {Tam}, S.~M. {Prakash}, J.~F. {Kennedy}, M.~S. {Ryerson},
  D.~{Lee}, and Y.~V. {Pant}, ``Design and implementation of a centralized
  system for autonomous unmanned aerial vehicle trajectory conflict
  resolution,'' in \emph{Proceedings of IEEE National Aerospace and Electronics
  Conference}, 2018.

\bibitem{kochenderfer2012next}
M.~J. Kochenderfer, J.~E. Holland, and J.~P. Chryssanthacopoulos,
  ``Next-generation airborne collision avoidance system,'' MIT-Lincoln
  Laboratory, Lexington, US, Tech. Rep., 2012.

\bibitem{ACASX}
J.-B. Jeannin, K.~Ghorbal, Y.~Kouskoulas, R.~Gardner, A.~Schmidt, E.~Zawadzki,
  and A.~Platzer, ``Formal verification of acas x, an industrial airborne
  collision avoidance system,'' in \emph{International Conference on Embedded
  Software}, Piscataway, NJ, USA, 2015.

\bibitem{UTMTCL4}
A.~Chakrabarty, C.~Ippolito, J.~Baculi, K.~Krishnakumar, and S.~Hening,
  ``Vehicle to vehicle (v2v) communication for collision avoidance for
  multi-copters flying in utm --tcl4,'' 01 2019.

\bibitem{ACASXu}
G.~{Manfredi} and Y.~{Jestin}, ``An introduction to acas xu and the challenges
  ahead,'' in \emph{2016 IEEE/AIAA 35th Digital Avionics Systems Conference
  (DASC)}, Sep. 2016, pp. 1--9.

\bibitem{DeCastro17}
J.~A. DeCastro, J.~Alonso-Mora, V.~Raman, and H.~Kress-Gazit, ``Collision-free
  reactive mission and motion planning for multi-robot systems,'' in
  \emph{Springer Proceedings in Advanced Robotics}, 2017.

\bibitem{Drona}
A.~Desai, I.~Saha, Y.~Jianqiao, S.~Qadeer, and S.~A. Seshia, ``Drona: A
  framework for safe distributed mobile robotics,'' in \emph{ACM/IEEE
  International Conference on Cyber-Physical Systems}, 2017.

\bibitem{AksarayCDC16}
D.~Aksaray, A.~Jones, Z.~Kong, M.~Schwager, and C.~Belta, ``Q-learning for
  robust satisfaction of signal temporal logic specifications,'' in \emph{IEEE
  Conference on Decision and Control}, 2016.

\bibitem{MaICUAS16}
X.~Ma, Z.~Jiao, and Z.~Wang, ``Decentralized prioritized motion planning for
  multiple autonomous uavs in 3d polygonal obstacle environments,'' in
  \emph{Inter. Conf. on Unmanned Aircraft Systems}, 2016.

\bibitem{4459804}
M.~{Kloetzer} and C.~{Belta}, ``A fully automated framework for control of
  linear systems from temporal logic specifications,'' \emph{IEEE Transactions
  on Automatic Control}, vol.~53, no.~1, pp. 287--297, Feb 2008.

\bibitem{1582935}
G.~E. {Fainekos}, H.~{Kress-Gazit}, and G.~J. {Pappas}, ``Hybrid controllers
  for path planning: A temporal logic approach,'' in \emph{Proce. of the 44th
  IEEE Conf. on Decision and Control}, Dec 2005, pp. 4885--4890.

\bibitem{1641832}
M.~{Kloetzer} and C.~{Belta}, ``Hierarchical abstractions for robotic swarms,''
  in \emph{Proc. of 2006 IEEE Inter. Conf. on Robotics and Automation}, May
  2006, pp. 952--957.

\bibitem{Raman14_MPCSTL}
V.~Raman, A.~Donze, M.~Maasoumy, R.~M. Murray, A.~Sangiovanni-Vincentelli, and
  S.~A. Seshia, ``Model predictive control with signal temporal logic
  specifications,'' in \emph{53rd IEEE Conf. on Decision and Control}, Dec
  2014, pp. 81--87.

\bibitem{Saha_acc16}
S.~Saha and A.~A. Julius, ``An milp approach for real-time optimal controller
  synthesis with metric temporal logic specifications,'' in \emph{Proceedings
  of the 2016 American Control Conference (ACC)}, 2016.

\bibitem{KaramanF11_LTLrouting}
S.~Karaman and E.~Frazzoli, ``Linear temporal logic vehicle routing with
  applications to multi-uav mission planning,'' \emph{International Journal of
  Robust and Nonlinear Control}, 2011.

\bibitem{pant2017smooth}
Y.~V. Pant, H.~Abbas, and R.~Mangharam, ``Smooth operator: Control using the
  smooth robustness of temporal logic,'' in \emph{Control Technology and
  Applications, 2017 IEEE Conf. on}.\hskip 1em plus 0.5em minus 0.4em\relax
  IEEE, 2017, pp. 1235--1240.

\bibitem{MalerN2004STL}
O.~Maler and D.~Nickovic, \emph{Monitoring Temporal Properties of Continuous
  Signals}.\hskip 1em plus 0.5em minus 0.4em\relax Springer Berlin Heidelberg,
  2004.

\bibitem{FainekosP09tcs}
G.~Fainekos and G.~Pappas, ``Robustness of temporal logic specifications for
  continuous-time signals,'' \emph{Theor. Computer Science}, 2009.

\bibitem{MuellerTRO15}
M.~W. Mueller, M.~Hehn, and R.~D\'Andrea, ``A computationally efficient motion
  primitive for quadrocopter trajectory generation,'' in \emph{IEEE
  Transactions on Robotics}, 2015.

\bibitem{PantAMNDM15_Anytime}
Y.~V. Pant, H.~Abbas, K.~Mohta, T.~X. Nghiem, J.~Devietti, and R.~Mangharam,
  ``Co-design of anytime computation and robust control,'' in \emph{2015 IEEE
  Real-Time Systems Symposium}, 2015.

\bibitem{hochreiter1997long}
S.~Hochreiter and J.~Schmidhuber, ``Long short-term memory,'' \emph{Neural
  computation}, vol.~9, no.~8, pp. 1735--1780, 1997.

\bibitem{gers2002learning}
F.~A. Gers, N.~N. Schraudolph, and J.~Schmidhuber, ``Learning precise timing
  with lstm recurrent networks,'' \emph{Journal of machine learning research},
  vol.~3, no. Aug, pp. 115--143, 2002.

\bibitem{lofberg2004yalmip}
J.~L{\"o}fberg, ``Yalmip: Toolbox for modeling and optim. in matlab.''

\bibitem{pastakia_faa_2015}
B.~Pastakia, J.~Won, R.~Sollenberger, D.~Aubuchon, S.~Entis, and L.~Thompson,
  ``{UAS Operational Assessment: Contingency Operations Simulation Report},''
  {Federal Aviation Administration}, Tech. Rep., 2015.

\end{thebibliography}
